\documentstyle[12pt,aasms4]{article}

\slugcomment{Draft \today}
\lefthead{}
\righthead{}

\def\wn{cm$^{-1}$}

\begin{document}
\title{The Second Measurement of Anisotropy in the Cosmic Microwave
Background Radiation\\
    at 0\fdg5 Scales\\
    near the Star Mu Pegasi}

\author{M. A. Lim\altaffilmark{1,3},
A. C. Clapp\altaffilmark{1,2,4},
M. J. Devlin\altaffilmark{1,2,5},
N. Figueiredo\altaffilmark{1,3,7,8},
J. O. Gundersen\altaffilmark{1,3},
S. Hanany\altaffilmark{1,2},
V. V. Hristov\altaffilmark{1,2},
A. E. Lange\altaffilmark{1,4},
P. M. Lubin\altaffilmark{1,3},
P. R. Meinhold\altaffilmark{1,3},
P. L. Richards\altaffilmark{1,2},
J. W. Staren\altaffilmark{1,3},
G. F. Smoot\altaffilmark{1,2,6},and
S. T. Tanaka\altaffilmark{1,2}}

\altaffiltext{1}{NSF Center for Particle Astrophysics,
	Berkeley, CA 94720.} 
\altaffiltext{2}{Physics Department,
	University of California at Berkeley,
	Berkeley, CA 94720} 
\altaffiltext{3}{Physics Department,
	University of California at Santa Barbara,
	Santa Barbara, CA 93106} 
\altaffiltext{4}{present address: Department of Physics,
	California Institute of Technology,
	Pasadena, CA 91125}
\altaffiltext{5}{present address: Department of Physics,
	Princeton University,
	Princeton, NJ 08544}
\altaffiltext{6}{Physics Division, Lawrence Berkeley Laboratory,
	Berkeley, CA 94720}
\altaffiltext{7} {
Departamento de F\'{\i}sica e Qu\'{\i}mica, Escola Federal de
Engenharia de 
Itajub\'{a}, 37500-000 Itajub\'{a}, MG, Brazil}
\altaffiltext{8} {
Divis\~{a}o de Astrof\'{\i}sica, Instituto Nacional de Pesquisas
Espaciais,
12201-970 S\~{a}o Jos\'{e} dos Campos, SP, Brazil} 
\setcounter{footnote}{0} 
\begin{abstract}
During the fifth flight of the Microwave Anisotropy Experiment
(MAX5), 
we revisited a region with significant dust emission near the star Mu
Pegasi.  A 3.5 cm$^{-1}$ low frequency channel has been added since
the previous measurement (\cite{mei93a}).
The data in each channel clearly show structure correlated with IRAS
100 \micron\ dust emission.  The spectrum of the structure in the 6,
9 and 14 cm$^{-1}$ channels is described by
$I_{\nu}\propto\nu^{\beta}B_{\nu}(T_{dust})$, where $\beta$ = 1.3 and
$T_{dust}$ = 19~K and $B_{\nu}$ is the Planck function.  However,
this model predicts a smaller amplitude in the 3.5 cm$^{-1}$
band than is observed. Considering only linear combinations of the
data independent of
the best fit foreground spectrum for the three lower channels, we
find an upper
limit to CMBR fluctuations of $\Delta T/T = \langle \frac{C_l~l(l+1)}{2\pi}\rangle^{\frac{1}{2}} \leq 1.3\times 10^{-5}$ at
the 95\% confidence level.  The result is for a flat band power
spectrum and does not include a 10\% uncertainty in calibration.  It
is consistent with our previous observation in the region.   
\end{abstract}

\keywords{cosmic microwave background --- cosmology:
	observations}

\section{Introduction}
Cosmic Microwave Background Radiation (CMBR) anisotropy measurements
provide a means of constraining various cosmological models.  Several
groups have reported measuring CMBR anisotropies at 0.5 to 1\arcdeg\
(\cite{che95,cla94,deb94,dev94,gun95,net95,ruh95}).  However,
disentangling the
primodial fluctuatations from foreground sources is problematic even
if the foreground is understood.  The third flight of MAX made an
observation in a medium constrast dust region near the star Mu Pegasi
and
measured anisotropy smaller than seen elsewhere in the same flight
(\cite{gun93,mei93a}).  In order to confirm this measurement, we
returned to the Mu Pegasi region with an additional low frequency
band centered
at 3.5 cm$^{-1}$.  

\section{Instrument}

MAX is an off-axis Gregorian telescope with a bolometric photometer
mounted on an attitude-controlled balloon platform.  The instrument
has been described extensively elsewhere
(\cite{fis92,als92,mei93b}).
The telescope has a 1~m off-axis parabolic primary with an elliptical
secondary which sinusoidally chops the beam in azimuth at
5.4~Hz with a peak-to-peak throw of 1\fdg4.  The chopped
signal is demodulated with a sine-wave lock-in reference.
The underfilled optics provide a 0\fdg5~FWHM beam.
An adiabatic
demagnetization refrigerator cools the single-pixel,
four-band photometer to 85~mK.  The four frequency bands are
centered at 3.5, 6, 9, and 14~cm$^{-1}$ with respective
fractional bandwidths 0.5, 0.5, 0.4, and 0.2.  To convert
measured antenna temperature differences to 2.726~K
thermodynamic temperature differences in each frequency band
multiply by 1.62, 2.50, 6.66, and 38.7, respectively.

\section{Observation}

The instrument was launched from the National Scientific
Balloon Facility in Palestine, Texas at 1.16~UT
June~20, 1994.  We observed CMBR anisotropies in three sky regions
near the stars HR5127, Phi Herculis, and Mu
Pegasi ( $\alpha$~=~22$^{\rm h}$49\fm7, $\delta$~=~24\fdg34\arcmin\
).
Tanaka et al. (1996) report on the observations at HR5127
and Phi Herculis.  This paper concerns the Mu Pegasi scan.

We observe with a constant velocity scan in azimuth of
$\pm$4\arcdeg\ relative to the pointing star.  The left-hand
lobe of the antenna pattern was coaligned with the field of
view of our CCD cameras and centered on Mu Pegasi.
Gyroscope drift was taken out every 400 seconds.  The
relative offset between the center of the chop and the
target star was 0\fdg55 in azimuth.

During the Mu Pegasi scan (7.22~UT to 7.76~UT) the gyro
malfunctioned and moved the chop center with a trajectory
tilted 10$\pm 1.5$ degrees from horizontal.  The orientation
of the gondola was still vertical.  We verified the
orientation and trajectory with the positions of stars in
the CCD camera field of view.  We did not observe in the same
orientation as in MAX3 and we do not expect the morphology
to be identical.  The other MAX 5 observations displayed no
significant tilt.

We calibrated the instrument before and after
the observation using a membrane transfer standard
(\cite{fis92}).  We observed Jupiter from 4.86~UT to 4.95~UT to
measure the
beam size and position and to confirm the membrane
calibration.  Using the best-fit beam size and the membrane
calibration, the derived temperature of Jupiter agrees with
Griffin et al. (1986) to within 10\%.  We assume a 10\%
uncertainty in calibration.  The calibration is such that a
chopped beam centered between sky regions with temperatures
$T_1$ and $T_2$ would yield $\Delta T = T_1 - T_2$ in the
absence of instrumental noise.

Anisotropy experiments are potentially susceptible to off-axis
response to local sources,
particularly the Earth, the balloon, and the moon.
The unchopped off-axis response in the 3.5 cm$^{-1}$
band is $\geq$ 70~dB below the on-axis response from
15\arcdeg\ to 25\arcdeg\ in elevation under the boresight.
We have not made comparable measurements of the chopped sidelobe
response in azimuth.
Mu Pegasi was $\sim$ 137\arcdeg\ away from the Moon
during the observation.

\section{Data Analysis}

\subsection{Data Reduction}

We remove transients due to cosmic rays using an algorithm
described by \cite{als92}.  This procedure excludes 
approximately 18\% of the data.  We demodulate the detector
output using the sinusoidal reference from the chopping
secondary to produce antenna temperature differences $\Delta
T_A$ on the sky.  This produces a data set in
phase and a data set 90\arcdeg\ out
of phase with the optical signal.  The noise averaged over
the observation gives respective CMBR sensitivities of 440,
240, 610, and 5100 $\mu K \sqrt{s}$ in the 3.5, 6, 9, and
14~cm$^{-1}$ bands.

The averages of the measured instrumental offsets in antenna
temperature were 0.6, 0.15, 1.4, and 2.8~mK in the 3.5, 6, 9, and 14
cm$^{-1}$ bands. We attribute this to chopped emissivity differences
on the primary mirror and chopped atmospheric emission.  The offset
drifts in the higher frequency bands with amplitudes of 0.7~mK and
1.0~mK in the 9 and 14 cm$^{-1}$ bands over a time scale of ~3
minutes.  Comparison of the first and second halves of the scan shows
that the signal is stable in the 3.5, 6, and 9 cm$^{-1}$ bands, but
not so in the 14 cm$^{-1}$ band.  The instability in the 14 cm$^{-1}$
band could be caused by sidelobe pickup or atmosphere. To increase
the stability, we subtract an offset and gradient, as in a ground
based observation, with a linear least squares fit to each pass going
from -4\arcdeg\ to +4\arcdeg\ or +4\arcdeg\ to -4\arcdeg. Each half
scan takes 72~s.    

For each observation we calculated the means and 1~$\sigma$
uncertainty of the antenna temperature differences for 29 pixels
separated by 17\arcmin\ on the sky.  Figure ~\ref{mpant} shows the
antenna temperature differences as a function of scan angle for the
Mu Pegasi scan.  There is significant structure ($\chi^2$~=~38, 86,
86, 79 for 27 DOF) that is well correlated ($R \gtrsim 0.5$) in all
channels of the in-phase data.  

\subsection{Foregrounds}
Possible astrophysical sources for the signal in the Mu Pegasi scan
are free-free or synchrotron radiation, interstellar dust (ISD)
emission, radio point sources, or CMBR. From the rising
spectrum in
$\Delta T_A$ in Figure~\ref{mpant} it
is clear that CMBR, free-free or
synchrotron radiation alone is not responsible for the signal. 
The latter two cases are also excluded by amplitude and
morphology arguments.
If we extrapolate the Haslam 408 MHz map (\cite{has87}) to
our frequencies using $\Delta T_A \propto \nu^{-2.1}$ for
free-free emission and $\Delta T_A \propto \nu^{-2.7}$
for synchrotron radiation and convolve with our chopped beam pattern,
we find that the former produces
$<10\%$ of the signal in the 3.5~cm$^{-1}$ channel and the latter
$<1\%$.  Furthermore, the morphology does not match
that of the data.   
An automated point source search \footnote{ The NASA/IPAC
   Extragalactic Database (NED) is operated by the Jet
   Propulsion Laboratory, California Institute of Technology,
under contract with the National Aeronautics and Space
Administration} has yielded no candidates
   within 90\arcmin\ that could produce a signal greater than
   $10~\mu K$.

Previous experience in this region leads us to expect ISD to be the
main contributor to our high frequency signal.  We convolved the IRAS
100 \micron\ maps with our chopped beam pattern and produced simulated scans.  We found the scale
factors that minimized the reduced $\chi^2$ from 100 \micron\ simulations to each
data channel separately and then normalized them to the 3.5 cm$^{-1}$
band.  The results are shown in Table \ref{specfits}.  The best fit
morphology and spectrum are superimposed over the data in Figure
~\ref{mpant}. If we consider the 6, 9, and 14 cm$^{-1}$ bands only,
these scale factors indicate a warm dust spectrum $ I_{\nu} \propto
\nu^{\beta} B_{\nu}(T_{dust})$, where $\beta$~=~1.3$_{-0.1}^{+0.2}$
and T$_{dust}$~=~19$_{-1}^{+1}$~K.  This is consistent with our
previous results (\cite{mei93a,fis95}).  However, the rise in
amplitude in the 3.5 cm$^{-1}$ band is not well explained by a warm
dust or warm and cold dust model.  

There are two possible causes for the rise in amplitude in the lowest
band.  One is a high frequency leak in the filters. Pre-flight
systematic tests with a thick grill high pass filter showed that high
frequency leakage above 20~\wn\ was less than 0.8\% of the total band
response to a 300K blackbody chopped relative to a 77K blackbody. Using
measured filter transmittances and the amplitude of dust fluctuations
in this sky region (Fischer et al. 1995), we calculated that maximum
modeled high frequency leakage of power from dust fluctuations
contributes less than $\sim$1\% of the expected inband power from CMB
fluctuations and less than 2\% of the observed structure. 
Another candidate is a
correlated low frequency component \footnote{\cite{kog95} report
correlation between HII and ISD at angular scales
$>7 \arcdeg$.  However, our cross-correlation coefficient 3.5
cm$^{-1}/\mathrm{IRAS}~100~\micron \mathrm{is}
30\pm6~\mathrm{mK(MJy/sr)}^{-1}$ which should be compared to $4.56
\pm 3.89~\mathrm{mK(MJy/sr)}^{-1}$ for $\mathrm{DMR}~90~\mathrm{GHz}/
\mathrm{DIRBE}~100~\micron$}.  However, fits of two component models
did not conclusively distinguish between the possibilities, such as
CMBR + ISD and HII + ISD.
  
We conclude the following about the foreground contaminant: The
correlation between the 14 cm$^{-1}$ band and the other bands
indicates a single foreground morphology.  Whatever the nature of the
foreground, the relative amplitudes in the bands are given in Table
\ref{specfits} column 2. Because of the excellent fit of the IRAS 100
\micron\ maps to the 14 cm$^{-1}$ channel, we assume that the ISD
dominates over any other possible high frequency contaminant.
    
\section{Discussion}

We analyze the three channels most sensitive to CMBR, 3.5, 6 and 9
cm$^{-1}$ for anisotropy in the presence of a single foreground
morphology with the spectrum found above.  We use maximum likelihood
methods assuming uniform prior to set
limits of the rms temperature fluctuation in the data, $Q \equiv
Q_{rms-PS} \equiv \langle Q_{rms} \rangle^{0.5}$
(\cite{smo92,wri94}). The
likelihood,
L, is given by 
\begin{equation}
\label{likelihood}
L \propto \frac{\exp(-\case{1}{2}
\mathbf{T}^{\mathnormal{T}}\mathbf{M}^{\mathnormal{-1}}\mathbf{T})}
{\sqrt{\det(\mathbf{M})}}
\end{equation}
where $\mathbf{T}$ is the data vector of all 29 bins and 3 channels
and $\mathbf{M}$ is the full covariance matrix for a flat band
power spectrum.  
 
We marginalize the data to account for the best fit foreground
spectrum given in Table \ref{specfits} and the offset and gradient
removal (\cite{dod94,bun94,bon91}). To do so, we construct a data
vector and covariance matrix, $\mathbf{T}^{\mathnormal{ind}} =
\mathbf{z}^{\mathnormal{T}}\mathbf{R}^{\mathnormal{T}}\mathbf{T}$ and
$\mathbf{M}^{\mathnormal{ind}} =
\mathbf{z}^{\mathnormal{T}}\mathbf{R}^{\mathnormal{T}}\mathbf{MRz}$,
where $\mathbf{z}$ and $\mathbf{R}$ account for both a single
foreground spectrum and offset and gradient removal respectively.
Using $\mathbf{M}^{\mathnormal{ind}}$ and
$\mathbf{T}^{\mathnormal{ind}}$ in equation (\ref{likelihood}) yields
an upper limit, $Q <$ 23 $\mu$K (95\% confidence level) or $\Delta
T/T~<~1.3 \times 10^{-5}$.  

When MAX3 Mu Pegasi is analyzed in a fashion similar to this paper,
we find $Q < 28~\mu$K (95\% confidence level) when marginalized for
the dust model in \cite{mei93a}.  The two data sets are consistent
with each other for similar analysis techniques.  Furthermore, the
MAX5 Mu Pegasi upper limit is consistent with the result from HR5127
($\Delta T/T = 1.2_{-0.3}^{+0.4} \times 10^{-5}$) although roughly so with Phi Herculis
($\Delta T/T = 1.9_{-0.4}^{+0.7} \times 10^{-5}$) which were also
observed in that flight.  

\section{Conclusion}
We have presented new results from a search for CMBR
anisotropy with high sensitivity at 0\fdg5 angular scales
near the star Mu Pegasi.  Free-free and synchrotron radiation are
excluded as 
the main source of signal on amplitude and spectral arguments. There
are no strong point sources in the field. The morphology of the
observed structure is consistent with known interstellar
dust but not the spectrum. The structure in
the 6, 9 and 14 cm$^{-1}$ channels is fit by a single dust model
power law $I_{\nu} \propto
\nu^{\beta} B_{\nu}(T_{dust})$, where $\beta = 1.3$, and
$T_{dust} = 19$~K.  We cannot rule out the possibility that the
structure
is a correlated combination of dust and CMBR or dust and free-free
radiation.  Linear
combinations of the data independent of the best fit
spectrum yield a $\Delta T/T~<~1.3 \times 10^{-5}$. (95\%
confidence level) The results are consistent
with our previous observation in the region.  These
data are available from the authors.
\acknowledgments

This work was supported by the National Science Foundation
through the Center for Particle Astrophysics (cooperative
agreement AST~91-20005), the National Aeronautics and Space
Administration under grants NAGW-1062 and FD-NAGW-1221, the
University of California, and previously the California
Space Institute. We would like to thank O.~Levy for his
assistance with the flight preparation and the MSAM team for
borrowed equipment during the flight campaign.  The IRAS map
used in the analysis was created by Skyview, maintained by
Goddard Space Flight Center.  N.F. was partially
supported by Conselho Nacional de Desenvolvimento Cient\'{\i}fico e 
Tecnol\'{o}gico, Brazil.

\clearpage
\begin{deluxetable}{crr}
\footnotesize
\tablecaption{Scale factors for fit to IRAS 100 \micron\ dust
morphology \label{specfits}}
\tablewidth{0pt}
\tablehead{
\colhead{Frequency (cm$^{-1}$)}
& \colhead{ $\frac{\Delta T_A^i}{\Delta T_A^{3.5}}$
\tablenotemark{a}}
& \colhead{Reduced $\chi^2$}
} 
\startdata
3.5 & 1.00 & 26/27	 \nl
6 & 0.55 & 56/27 \nl
9 & 0.74 & 43/27 \nl
14 & 1.00 & 28.5/27 \nl
\enddata
\tablenotetext{a}{ These are the ratios of the differential antenna
temperatures for the best fit spectrum}
\end{deluxetable}

\clearpage

\figcaption{ Antenna temperature differences plotted as a function of
scan
angle. Superimposed over the data are IRAS 100 \micron\ dust
morphologies
scaled for different spectra.  The dashed line is for a spectrum with
$\beta~=~1.3$ and $T_{dust}~=~19~K$. The solid line is for a similar
morphology
but with the amplitudes chosen to minimize $\chi^2$. 
\label{mpant} }
  
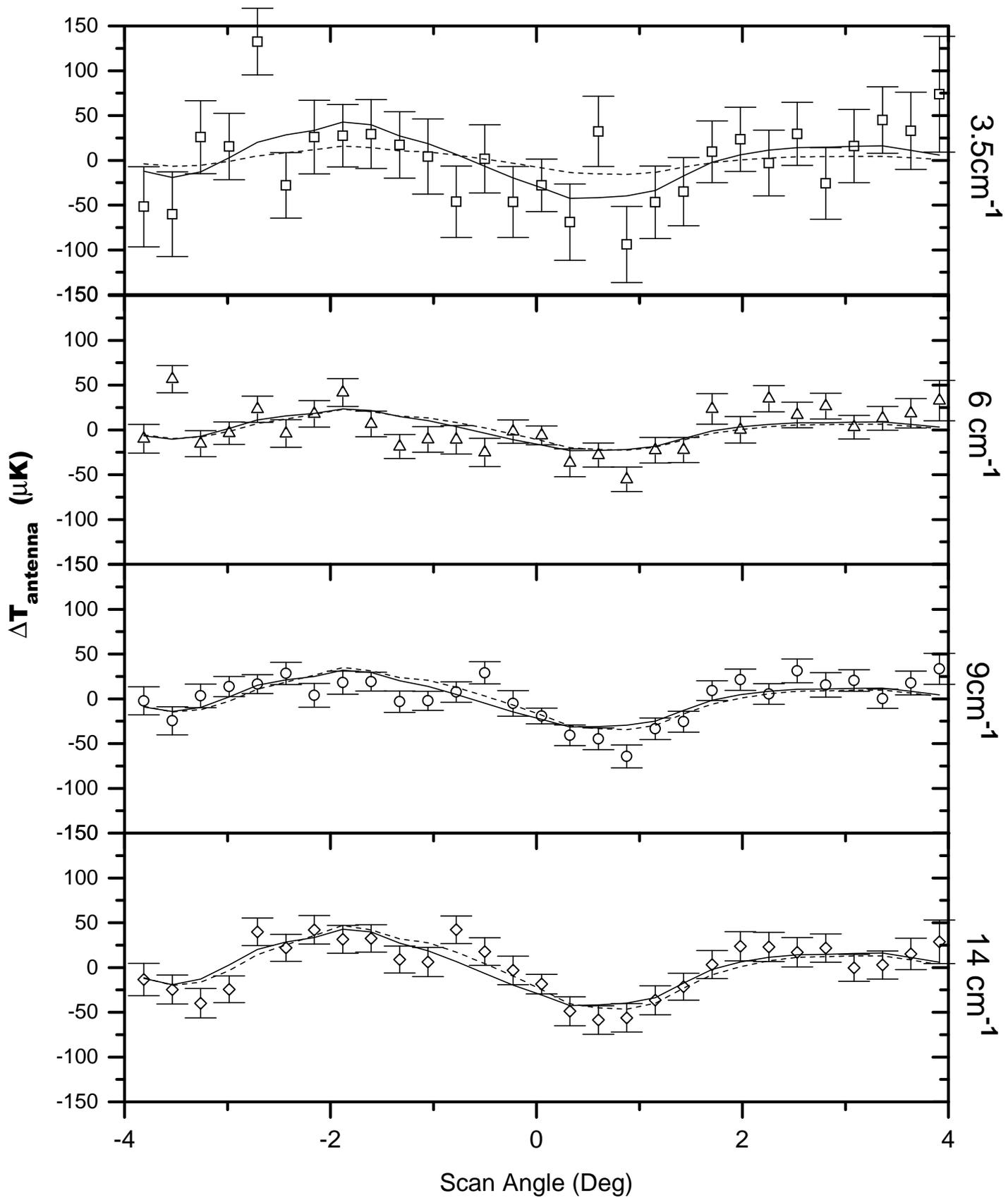
\end{document}